# Classical Heisenberg Hamiltonian Solution of Oriented Spinel Ferrimagnetic Thin Films

P. Samarasekara

Department of Physics, University of Ruhuna, Matara, Sri Lanka.

**Abstract**

The classical Heisenberg Hamiltonian was solved for oriented spinel thin and thick cubic ferrites. The dipole matrix of complicated cubic cell could be simplified into the form of dipole Matrix of simple cubic cells. This study was confined only to the highly oriented thin films of ferrite. The variation of total energy of Nickel ferrite thin films with angle and number of layers was investigated. Also the change of energy with stress induced anisotropy for Nickel ferrite films with N=5 and 1000 has been studied. Films with the magnetic moments ratio 1.86 can be easily oriented in $\theta=90^0$ direction when N is greater than 400. Although this simulation was performed only for $\frac{J}{\omega}=100, \frac{\sum_{m=1}^{N} D_m^{(2)}}{\omega}=10, \frac{H_{in}}{\omega}=\frac{H_{out}}{\omega}=0, \frac{K_s}{\omega}=5 \text{ and } \frac{\sum_{m=1}^{N} D_m^{(4)}}{\omega}=5$ as an example, these equations can be applied for any value of $\frac{J}{\omega}, \frac{\sum_{m=1}^{N} D_m^{(2)}}{\omega}, \frac{H_{in}}{\omega}, \frac{H_{out}}{\omega}, \frac{K_s}{\omega}$ and $\frac{\sum_{m=1}^{N} D_m^{(4)}}{\omega}$.

## 1. Introduction:

For the first time dipole matrix and the total energy of the cubic spinel ferrites were calculated using classical model of Heisenberg Hamiltonian in detail. All the relevant energy terms such as spin exchange energy, dipole energy, second and fourth order anisotropy terms, interaction with magnetic field and stress induced anisotropy in Heisenberg Hamiltonian were taken into account. The spin exchange interaction energy and dipole interaction have been calculated only between two nearest spin layers and within same spin plane. These equations derived here can be applied for spinel ferrites such as $Fe_3O_4$, $NiFe_2O_4$ and $ZnFe_2O_4$ only. But these equations can not be applied for ferrites such as Lithium ferrite.

The structure of spinel ferrites with the position of octahedral and tetrahedral sites is given in detail in some early report [1-5]. Although there are many filled and vacant octahedral and tetrahedral sites in cubic spinel cell [1], only the occupied octahedral and tetrahedral sited were used for the calculation in this report. Only few previous reports could be found on the theoretical works of ferrites [6-9]. The solution of Heisenberg ferrites consist of spin exchange interaction term only has been found earlier by means of the retarded Green function equations [6].

## 2. The model

The Hamiltonian in Heisenberg model can be written as following for a film.

$$H = -J\sum_{m,n}\vec{S}_m.\vec{S}_n + \omega\sum_{m\neq n}(\frac{\vec{S}_m.\vec{S}_n}{r_{mn}^3} - \frac{3(\vec{S}_m.\vec{r}_{mn})(\vec{r}_{mn}.\vec{S}_n)}{r_{mn}^5}) - \sum_m D_{\lambda_m}^{(2)}(S_m^z)^2 - \sum_m D_{\lambda_m}^{(4)}(S_m^z)^4$$

$$-\sum_m \vec{H}..\vec{S}_m - \sum_m K_s Sin2\theta_m \quad \textbf{(1)}$$

Here θ is the angle between local magnetization (M) and the stress. Within a single domain, M is parallel to the spin. If stress is applied normal to the film plane, then $\theta_m$ is the angle between the normal to the film plane and the local spin. Here the last term indicates the change of magnetic energy under the influence of a stress. $K_s$ depends on the product of magnetostriction coefficient ($\lambda_s$) and the stress ($\sigma$). $K_s$ can be positive or negative depending on the type of stress whether it is compressive or tensile. Integer m and n denote the indices of planes, and they vary from 1 to N for a film with N number of layers. First, second, third and fourth terms represent the spin exchange interaction, magnetic dipole interaction, second order anisotropy and fourth order anisotropy, respectively. Here $\vec{S}_i$ is a spin vector at point $\vec{r}_i$ in layer $\lambda_i$. Therefore the ground state energy will be calculated per spin with Z-axis normal to film plane. $\vec{H}$ is the external magnetic field with the effective magnetic moment µ of the spins incorporated.

The spin exchange interaction energy is negative and positive for parallel and antiparallel spins, respectively. Hence spin exchange interaction energy constant (J) is positive and negative for parallel (Ferromagnetic) and antiparallel (ferrites or antiferromagnetic) spin arrangements, respectively. Similarly dipole interaction energy is positive and negative for parallel and antiparallel spin arrangements. In this report, the spin structure of $AFe_2O_4$ spinel ferrite cell described by Kurt et al will be used [1]. In this considered spinel ferrite cubic cell, all spins in one spin layer are produced by either Iron or other metal (A) ions, but spins in two consecutive spin layers are produced by Iron and metal (A) ions alternatively. Therefore, all the spins in one layer are parallel, and spins in consecutive layers are antiparallel. Although dipole interaction energy and J are positive within one spin layer, both of them are negative between two nearest spin layers. Only the interaction between two nearest layers has been considered for these calculations.

## 3. Results and discussion

The length of one side of the cubic cell was taken as "a". The spin exchange or dipole interaction between two spins with separation less than "a" was taken into account. Eight spin layers with separation a/8 in the cubic unit cell were considered as given in Kurt et al [1]. The number (Z) of metal (A) and Iron ions with separation less than "a" in each layer and in between two nearest spin layers is given in table 3. The spin with magnitude s will be given as s(0, $\sin\theta_\mu$, $\cos\theta_\mu$). The dipole interaction energy between two spins is given by

$$E = \omega \vec{S}_i . W(r_{ij}) . \vec{S}_j \quad (2)$$

Here 
$$W(r) = \frac{1}{r^3} \begin{pmatrix} 1-3\hat{r}_x^2 & -3\hat{r}_y\hat{r}_x & -3\hat{r}_z\hat{r}_x \\ -3\hat{r}_x\hat{r}_y & 1-3\hat{r}_y^2 & -3\hat{r}_z\hat{r}_y \\ -3\hat{r}_x\hat{r}_z & -3\hat{r}_y\hat{r}_z & 1-3\hat{r}_z^2 \end{pmatrix} \quad (3)$$

and $\omega = \dfrac{\mu_0 \mu^2}{4\pi a^3}$

The spins of A and Fe ions are given as 1 and p, respectively. For an example, the ratio between spins in Nickel ferrite ($NiFe_2O_4$) can be given as p=2.5. The matrix elements calculated within each in spin layer and in between two nearest spin layers are also given in table 3. A film with (001) orientation of spinel ferrite cell has been considered. As an example, the calculations of some matrix elements are given below. In layer one (bottom layer of spinel cell), five metal ions occupy $A_1$(0, 0, 0), $A_2$(1, 0, 0), $A_3$(0, 1, 0), $A_4$(1, 1, 0) and $A_5$(0.5, 0.5, 0) sites [1]. Because the interactions between spins with separation less than "a" have been considered, only the $A_5A_1$, $A_5A_2$, $A_5A_3$ and $A_5A_4$ spin interactions have been taken into account. For these spin interactions, individual and total matrix elements are given in table 1. The spins of Fe ions in the second spin layer located at 0.125 above the bottom layer of spinel cell occupy $Fe_1$(0.125, 0.625, 0.125), $Fe_2$(0.375, 0.875, 0.125), $Fe_3$(0.625, 0.125, 0.125) and $Fe_4$(0.875, 0.375, 0.125) sites [1]. Because the separations between following interactions are less than "a", the spin interactions between $A_3Fe_1$, $A_3Fe_2$, $A_2Fe_3$, $A_2Fe_4$, $A_5Fe_1$, $A_5Fe_2$, $A_5Fe_3$ and $A_5Fe_4$ were considered for this calculations. The individual and total dipole matrix elements of these interactions are given in table 2. Similarly the total dipole matrix elements calculated for other spin layers are given in table3.

The spin exchange interaction energy for nearest spins in one layer and spins between nearest layers with separation less than "a" can be given as below for one unit cell using the nearest spin neighbors given in table 3. J is assumed to be a constant for all spin layers throughout the whole film.

$E^{Ex}_{unit\ cell} = -10J + 80Jp - 22Jp^2$

For a thin film with thickness Na (height of N cubic cells),

Total spin exchange interaction energy = $E^{Ex}_{Total} = 2J(-5N + 40Np - 11p^2N - 4p)$ **(4)**

For all A type spins given in table 3, the dipole interaction energy can be given as following by using equation 2.

$$E^{dipole}_A = \omega \begin{pmatrix} 0 & \sin\theta_\mu & \cos\theta_\mu \end{pmatrix}(-28.2842)\begin{pmatrix} \frac{1}{2} & 0 & 0 \\ 0 & \frac{1}{2} & 0 \\ 0 & 0 & -1 \end{pmatrix}\begin{pmatrix} 0 \\ \sin\theta_\nu \\ \cos\theta_\nu \end{pmatrix}$$

Here 28.2842 is the addition of $W_{33}$ matrix elements of all A type rows given in table 3. Also this dipole matrix is similar to that of a highly symmetric cubic cell [10]. Within one ferrite unit cell, all the spins are either parallel or antiparallel to each other due to the super exchange interaction between spins. Therefore, the angle $\theta_\mu=\theta_\nu=\theta$ within unit cell will deduce above equation to following equation.

$$E^{dipole}{}_A = -\omega 28.2842(\frac{1}{4}+\frac{3}{4}\cos 2\theta)$$

Similarly for all Fe layers and A-Fe nearest layer interactions,

$$E^{dipole}{}_B = -\omega p^2 257(\frac{1}{4}+\frac{3}{4}\cos 2\theta)$$

$$E^{dipole}{}_{AB} = p\omega 576.3464(\frac{1}{4}+\frac{3}{4}\cos 2\theta)$$

The dipole interaction energy of a unit cell= $E^{dipole}{}_{unitcell} = E^{dipole}{}_A + E^{dipole}{}_B + E^{dipole}{}_{AB}$

If the film is highly oriented the angle $\theta$ remains same throughout the whole film.

For a thin film of thickness Na,

Total dipole interaction energy= $E^{dipole}{}_{Total} = NE^{dipole}{}_{unitcell}$

$$E^{dipole}{}_{Total} = \omega N(\frac{1}{4}+\frac{3}{4}\cos 2\theta)(-28.2842 - 257p^2 + 576.3464p) \qquad (5)$$

This equation is valid only for large N. The 5$^{th}$ term of equation 1 can be given as,

$$\sum_m \vec{H}.\vec{S}_m = 4N(H_{in}\sin\theta + H_{out}\cos\theta)(1-p) \qquad (6)$$

Therefore, from equation 1, 4, 5 and 6 the total energy can be given as,

$$E(\theta) = 2J(-5N + 40Np - 11p^2N - 4p)$$

$$+ \omega N(\frac{1}{4}+\frac{3}{4}\cos 2\theta)(-28.2842 - 257p^2 + 576.3464p)$$

$$-\cos^2\theta \sum_{m=1}^{N} D_m^{(2)} - \cos^4\theta \sum_{m=1}^{N} D_m^{(4)} - 4N(1-p)(H_{in}\sin\theta + H_{out}\cos\theta + K_s \sin 2\theta) \qquad (7)$$

Using $\frac{J}{\omega} = 100, H_{in} = 0, \theta = 90^0$ and $\frac{\partial E}{\partial p} = 0$ for minimum energy,

$$p = 1.86 - \frac{0.19}{N}$$

The graph between p and N is given in figure 1. The number of layers corresponding for preferred perpendicular orientation can be found for different ferrites using this graph. For ferrites with p=1.86, the films with N>400 can be easily oriented in $\theta=90^0$ direction. For Nickel ferrite, p=2.5

$$E(\theta) = 2J(26.25N - 10) - 48.415\omega N(1 + 3\cos 2\theta)$$

$$- \cos^2 \theta \sum_{m=1}^{N} D_m^{(2)} - \cos^4 \theta \sum_{m=1}^{N} D_m^{(4)} + 6N(H_{in} \sin\theta + H_{out} \cos\theta + K_s \sin 2\theta) \quad \textbf{(8)}$$

When $\dfrac{J}{\omega} = 100, \dfrac{\sum_{m=1}^{N} D_m^{(2)}}{\omega} = 10, \dfrac{H_{in}}{\omega} = \dfrac{H_{out}}{\omega} = 0, \dfrac{K_s}{\omega} = 5$ and $\dfrac{\sum_{m=1}^{N} D_m^{(4)}}{\omega} = 5$

$$\frac{E(\theta)}{\omega} = 5201.6N - 2000 - 145.25N\cos 2\theta - 10\cos^2\theta - 5\cos^4\theta + 30N\sin 2\theta$$

The 3-D graph of $\dfrac{E(\theta)}{\omega}$ versus N and θ is given in figure 2. When the number of layers increases, the energy gradually increases with some sinusoidal variation. But for one value of N, the energy remains constant.

When N=5 (thin) and $\dfrac{K_s}{\omega}$ is a variable from equation 8,

$$\frac{E(\theta)}{\omega} = 247343 - 1462.6\cos^2\theta - 5\cos^4\theta + 30\frac{K_s}{\omega}\sin 2\theta$$

Here the other constants given above have been used. The 3-D plot of $\dfrac{E(\theta)}{\omega}$ versus θ and $\dfrac{K_s}{\omega}$ is given in figure 3. Because the energy indicates minimums at some stress values, the value of stress corresponding to different orientations with minimum energy can be obtained using this graph.

When N=1000 (thick) and $\frac{K_s}{\omega}$ is a variable,

$$\frac{E(\theta)}{\omega} = 5344830 - 290500\cos^2\theta - 5\cos^4\theta + 6000\frac{K_s}{\omega}\sin 2\theta$$

The 3-D plot of $\frac{E(\theta)}{\omega}$ versus $\theta$ and $\frac{K_s}{\omega}$ for N=1000 is given in figure 4. According to graph 3 and 4, the stress corresponding to minimum energy increases with number layers at lower angles. But at higher angles, this stress corresponding to minimum energy does not depend on number of layers. Also according to these two graphs, the maximum energy increases with the number of layers.

For $Fe_3O_4$ and $ZnFe_2O_4$, the ratio p=1.25 and 1, respectively. Above simulation can be repeated for theses ferrites too by using equation 7.

**4. Conclusion**

The dipole matrix of this complicated spinel cubic cell could be simplified into the form of dipole Matrix of simple cubic cells. Films with the magnetic moments ratio 1.86 can be easily oriented in $\theta=90^0$ direction when N is greater than 400. The total energy of Nickel ferrite thin films gradually increases with number of layers for the values used in this simulation. Also the energy indicates some minimum values at some stress values implying that the film can be easily oriented in some directions under the influence of some particular applied stress. Also this stress corresponding to minimum energy varies with number of layers at lower angles. This simulation can be similarly applied for any value of $\dfrac{J}{\omega}, \dfrac{\sum_{m=1}^{N} D_m^{(2)}}{\omega}, \dfrac{H_{in}}{\omega}, \dfrac{H_{out}}{\omega}, \dfrac{K_s}{\omega}$ and $\dfrac{\sum_{m=1}^{N} D_m^{(4)}}{\omega}$.

**References:**

1. Kurt E. Sickafus, John M. Wills and Norman W. Grimes, J. Am.Ceram. Soc. **82(12)**, 3279 (1999)

2. I.S. Ahmed Farag, M.A. Ahmed, S.M. Hammad and A.M. Moustafa Egypt, J. Sol. **24(2)**, 215 (2001)

3. V. Kahlenberg, C.S.J. Shaw and J.B. Parise, Am.Mineralogist **86**, 1477 (2001)

4. I.S. Ahmed Farag, M.A. Ahmed, S.M. Hammad and A.M. Moustafa, Cryst. Res. Technol. **36**, 85 (2001)

5. Z. John Zhang, Zhong L. Wang, Bryan C. Chakoumakos and Jin S. Yin, J. Am. Chem. Soc. **I20**, 1800, (1998)

6. Ze-Nong Ding, D.L. Lin and Libin Lin, Chinese J. Phys. **31(3)**, 431 (1993)

7. D. H. Hung, I. Harada and O. Nagai, Phys. Lett. **A53**, 157 (1975)

8. H. Zheng and D.L. Lin, Phys Rev. **B37**, 9615 (1988)

9. S.T. Dai and Z.Y. Li, Phys. Lett. **A146**, 50 (1990)

10. K.D. Usadel and A. Hucht: Phys. Rev. B **66**, 024419-1 (2002)

**Figure and Table Captions**

Table 1. The individual and total dipole matrix elements of the bottom layer of spinel cell

Table2. The individual and total dipole matrix elements for the interaction between first and second layer of spinel cell

Table 3. The number of nearest neighbors and matrix elements of dipole tensor for each layer and two nearest layers

Figure 1. Graph between p and N at minimum energy for perpendicular orientation

Figure 2. 3-D graph of $\frac{E(\theta)}{\omega}$ versus N and θ for Nickel ferrite

Figure 3. 3-D plot of $\frac{E(\theta)}{\omega}$ versus θ and $\frac{K_s}{\omega}$ for Nickel ferrite with N=5

Figure 4. 3-D plot of $\frac{E(\theta)}{\omega}$ versus θ and $\frac{K_s}{\omega}$ for Nickel ferrite with N=1000

|       | $W_{11}$ | $W_{12}=W_{21}$ | $W_{13}=W_{31}$ | $W_{22}$ | $W_{23}=W_{32}$ | $W_{33}$ |
|-------|----------|-----------------|-----------------|----------|-----------------|----------|
| $A_5A_1$ | -1.41421 | -4.24264 | 0 | -1.41421 | 0 | 2.828427 |
| $A_3A_5$ | -1.41421 | 4.242641 | 0 | -1.41421 | 0 | 2.828427 |
| $A_2A_5$ | -1.41421 | 4.242641 | 0 | -1.41421 | 0 | 2.828427 |
| $A_4A_5$ | -1.41421 | -4.24264 | 0 | -1.41421 | 0 | 2.828427 |
| Total | -5.65685 | 0 | 0 | -5.65685 | 0 | 11.31371 |

Table 1

|  | $W_{11}$ | $W_{12}=W_{21}$ | $W_{13}=W_{31}$ | $W_{22}$ | $W_{23}=W_{32}$ | $W_{33}$ |
|---|---|---|---|---|---|---|
| $A_2Fe_3$ | -20.4131 | 11.48235 | 11.48235 | 10.20653 | -3.82745 | 10.20653 |
| $A_2Fe_4$ | 10.20653 | 11.48235 | 3.82745 | -20.4131 | -11.4823 | 10.20653 |
| $A_5Fe_3$ | 10.20653 | 11.48235 | -3.82745 | -20.4131 | 11.48235 | 10.20653 |
| $A_5Fe_4$ | -20.4131 | 11.48235 | -11.4823 | 10.20653 | 3.82745 | 10.20653 |
| $A_3Fe_1$ | 10.20653 | 11.48235 | -3.82745 | -20.4131 | 11.48235 | 10.20653 |
| $A_3Fe_2$ | -20.4131 | 11.48235 | -11.4823 | 10.20653 | 3.82745 | 10.20653 |
| $A_5Fe_1$ | -20.4131 | 11.48235 | 11.48235 | 10.20653 | -3.82745 | 10.20653 |
| $A_5Fe_2$ | 10.20653 | 11.48235 | 3.82745 | -20.4131 | -11.4823 | 10.20653 |
| Total | -40.8261 | 91.8588 | 0 | -40.8261 | 0 | 81.65226 |

Table 2

| Layer | Z | $W_{11}$ | $W_{12}=W_{21}$ | $W_{13}=W_{31}$ | $W_{22}$ | $W_{23}=W_{32}$ | $W_{33}$ |
|---|---|---|---|---|---|---|---|
| 1 Metal | 4 | -5.65685 | 0 | 0 | -5.65685 | 0 | 11.31371 |
| 1 and 2 | 8 | -40.8261 | 91.8588 | 0 | -40.8261 | 0 | 81.65226 |
| 2 Fe | 6 | -27.4797 | -55.754 | 0 | -27.4797 | 0 | 54.9594 |
| 2 and 3 | 8 | -26.9009 | 37.81957 | 0 | -26.9009 | 0 | 53.80186 |
| 3 Metal | 1 | -1.41421 | -4.24264 | 0 | -1.41421 | 0 | 2.828427 |
| 3 and 4 | 8 | -26.9009 | -37.8196 | 0 | -26.9009 | 0 | 53.80186 |
| 4 Fe | 5 | -36.7696 | 110.3087 | 0 | -36.7696 | 0 | 73.53911 |
| 4 and 5 | 16 | -49.4587 | -78.9235 | 0 | -49.4587 | 0 | 98.91731 |
| 5 Metal | 4 | -5.65685 | 0 | 0 | -5.65685 | 0 | 11.31371 |
| 5 and 6 | 16 | -49.4587 | 78.92348 | 0 | -49.4587 | 0 | 98.91731 |
| 6 Fe | 5 | -36.7696 | -110.309 | 0 | -36.7696 | 0 | 73.53911 |
| 6 and 7 | 8 | -26.9009 | 37.81957 | 0 | -26.9009 | 0 | 53.80186 |
| 7 Metal | 1 | -1.41421 | 4.242641 | 0 | -1.41421 | 0 | 2.828427 |
| 7 and 8 | 8 | -26.9009 | -37.8196 | 0 | -26.9009 | 0 | 53.80186 |
| 8 Fe | 6 | -27.4797 | 55.75403 | 0 | -27.4797 | 0 | 54.9594 |
| 8 and 9 | 8 | -40.8261 | -91.8588 | 0 | -40.8261 | 0 | 81.65226 |

Table 3

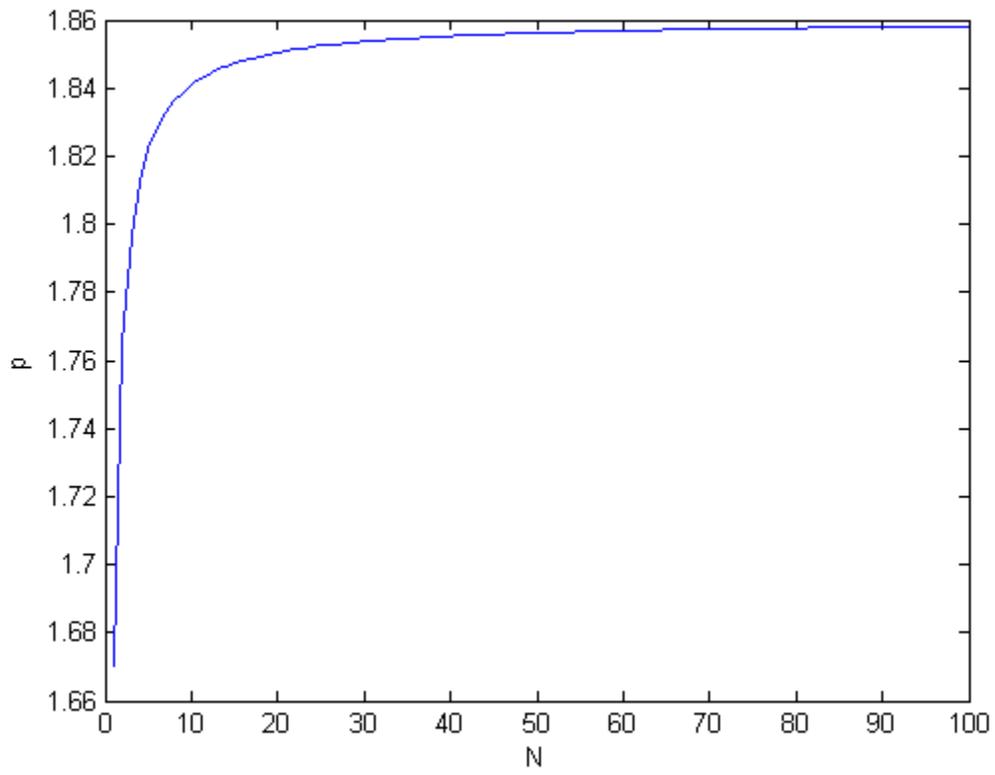

Figure 1

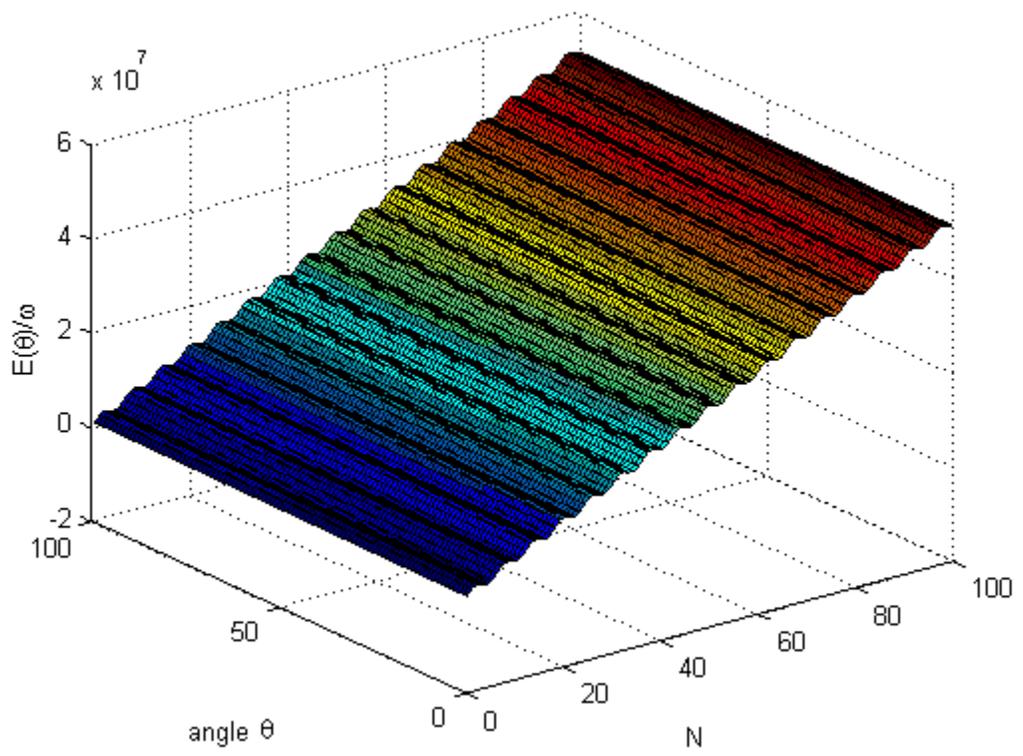

Figure 2

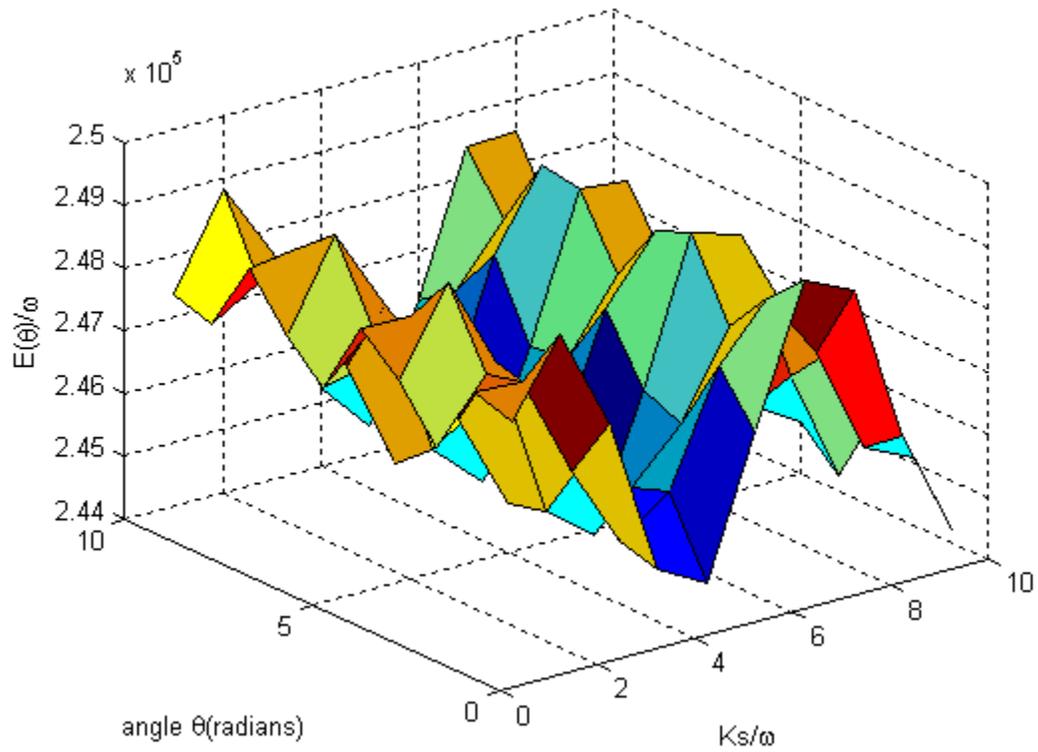

Figure 3

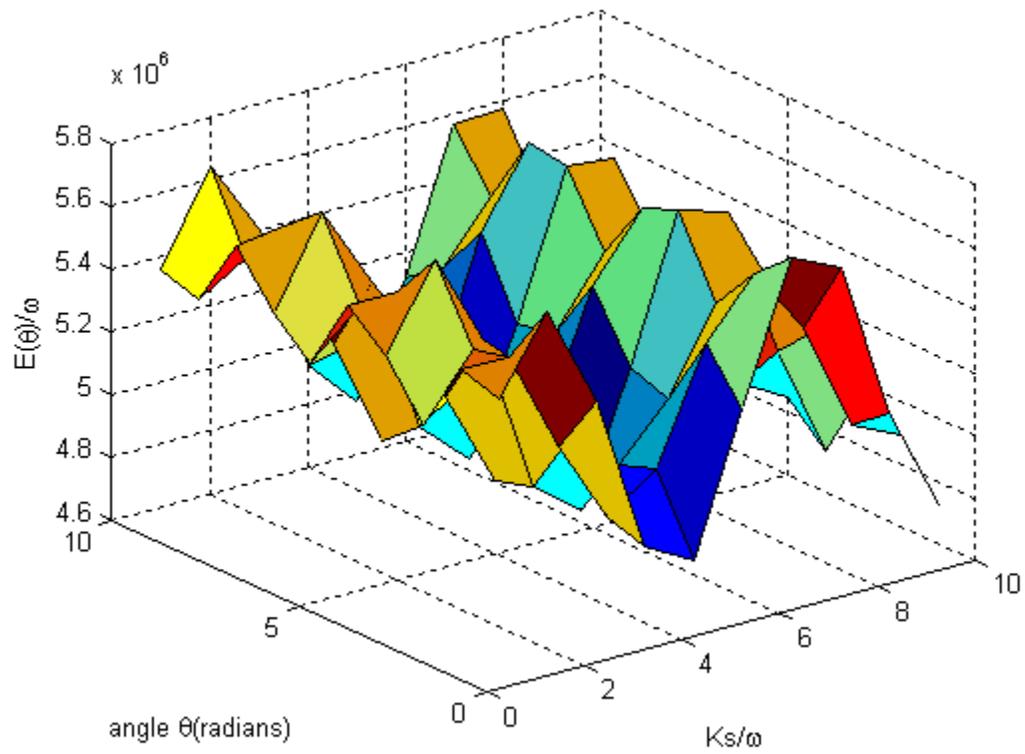

Figure 4